# Investigation of Io's Auroral Hiss Emissions Due To Its Motion in Jupiter's Magnetosphere

ASEG-2012

M.H.Moghimi[1]
*M.Sc. in Aerospace Engineering, Sharif University of Technology, Tehran, 1581955711, Iran*

**Abstract:** The left-hand side of the auroral hiss emission observed by Galileo has a frequency time shaped very similar to the funnel shape observed in the earth's auroral region. This close similarity indicates that we can use the whistler-mode propagation near resonance cone to locate the emission source. In this paper the general characteristic of the whistler mode are discussed. Then the position of the emission source has been investigated using a geometry method that takes into account the Galileo's trajectory. Initially it is assumed the source is a point. Then the possibility of sheet source aligned along the magnetic field lines which are tangent to the surface of Io is investigated. Both of two types of sources show that the whistler mode radiation originates very close to the surface of the Io.

## Nomenclature

| | | |
|---|---|---|
| $\boldsymbol{B}$ | = | magnetic field |
| $\overset{\leftrightarrow}{\sigma}$ | = | conductivity tensor |
| $\overset{\leftrightarrow}{K}$ | = | dielectric tensor |
| $\boldsymbol{k}$ | = | wave propagation vector |
| $\boldsymbol{J}$ | = | current density |
| $\boldsymbol{E}$ | = | electric field |
| $\omega_{c\sigma}, \omega_{p\sigma}$ | = | cyclotron frequency, and plasma frequency of particle $\sigma$ respectively |
| $\omega$ | = | wave frequency |
| c | = | phase velocity of the light waves |
| $\boldsymbol{n}$ | = | refractive index, $\left(\frac{c\boldsymbol{k}}{\omega}\right)$ |
| $\varepsilon_0, \mu_0$ | = | permittivity of vacuum, and permeability of vacuum respectively |
| $\rho$ | = | charge density |
| $q_\sigma, m_\sigma, n_\sigma$ = | | particle's charge, particle's mass, and particle's density respectively |
| $e$ | = | electron charge |
| $\boldsymbol{V}_c, \boldsymbol{V}_\sigma, \boldsymbol{V}_A$ = | | conductor's velocity, particle's velocity, and Alfven velocity respectively |

---

[1] M.Sc. in Aerospace Engineering, Aerospace Engineering Department, m_moghimi@alum.sharif.edu.

## I. Introduction

Auroral hiss is a type of whistler mode radio emission which frequently occurs in high latitude regions of planetary magnetospheres and also observed by Galileo spacecraft over the Jupiter's moon Io during a flyby of Io in October 16, 2001. This emission was first discovered using ground based instruments, [1] which detected a very low frequency broadband emission in association with aurora. The first satellite observations of auroral hiss were made at the earth in the 1960s, [3,4,5,6] and similar emissions have since been detected at other planets in the solar system, including Jupiter, Jupiter's moon Io, and Saturn. Auroral hiss is known to propagate in the whistler mode, as the frequency is always above the local proton cyclotron frequency and below both the electron cyclotron frequency and electron plasma frequency. Whistler mode wave are the only plasma waves that can propagate in this range of frequencies. On a frequency-time spectrogram, this emission displays a characteristic funnel shape with a V-shaped low frequency cutoff as shown by Gurnett. [4] The emission's V-shaped nature is explained by emission from localized source propagation near the resonance cone. [7,8,9] The resonance cone is defined as a cone of angles with respect to the magnetic field where the refractive index goes to infinity, and denotes the region around the magnetic field line in which the emission is restricted to propagate. The resonance cone also imposes a restriction on the group velocity, the propagation of wave energy, to a region of angles around the field line. This boundary is a property of the index of refraction, $n(\theta)$, where $\theta$ is the angle between the propagation vector and the central field line. This is shown in Fig. 1. As the refractive vector deviates further from the magnetic field direction, it approaches an angle where the index of refraction asymptotes to infinity. This angle is defined as $\theta_{res}$. From this angle the limiting group velocity angle $\psi$ could be defined. It can be shown that the group velocity angle is perpendicular to the index of refraction surface. [10] Then it can be written $\psi = 90^o - \theta_{res}$. This condition, however, is not uniform for all components of the emission; because the resonance cone angle is frequency dependent. The lower frequencies will have a smaller resonance cone angle than higher frequencies. In addition, this emission has an upper cutoff, as the resonance cone angle goes to $\theta_{res}$ at either the plasma frequency or cyclotron frequency. These results imply that as a spacecraft flies through a region of whistler mode emission propagating along the resonance cone it will encounter higher frequencies first, and detect lower frequencies only near the midpoint of its pass through this region. This emission will also be bounded on the high frequency end by either the plasma frequency or cyclotron frequency whichever is lower.

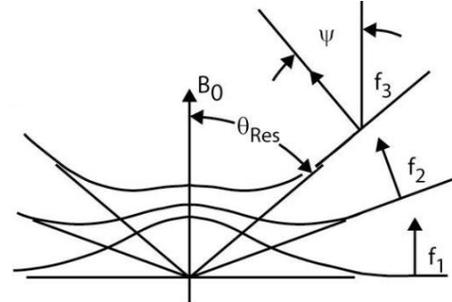

**Figure 1. Propagation of auroral hiss emission along the resonance cone from Ref. 2.**

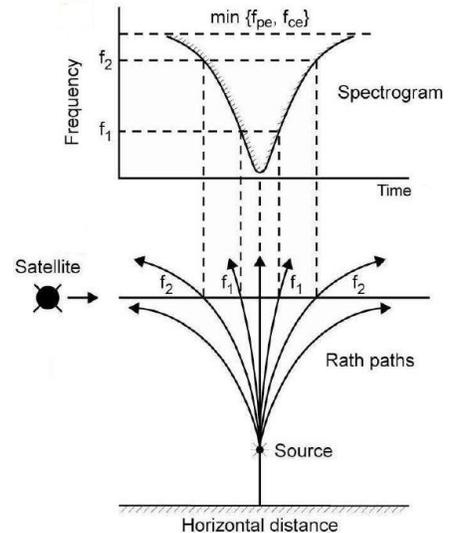

**Figure 2. Frequency Dependent Propagation and its effect on how data appear on a spectrogram from Ref. 2.**

This characteristic is shown in Fig. 2. As one can see in Fig. 2, this emission will display a characteristic funnel shape when plotted on a frequency versus time spectrogram. It is clear that the detection of such emissions is dependent upon the location of spacecraft as it orbits the planet. In the following sections first the observations and the unipolar inductor model will be reviewed. Then after presenting the whistler mode generation and propagation, both point source and cylindrical sheet geometry model will be discussed.

## II. Observations and the Unipolar Inductor Model

In this section we proceed by reviewing the frequency-time spectrogram taken by Galileo and then will discuss current generation of conductor which pass through a plasma environment in brief.

### A. The Galileo's Observations

The Galileo spacecraft which was placed in the orbit around Jupiter on December 7, 1995, has been carrying out a series of close flybys of the four Galilean satellites. The spacecraft trajectory relative to Io from 00:50:00 UT to 02:10:00 UT on October, 16, 2001 is shown in Fig. 3. In this figure the coordinate system is used with the +Z axis aligned parallel to Jupiter's rotational axis and the +X axis aligned parallel to the nominal co-rotational plasma flow induced by Jupiter's rotation. The +Y axis completes the usual right-handed coordinate system. As can be seen, the spacecraft passed over the south pole of Io with a closest approach at a radial distance of 1.098 $R_{Io}$ at 01:23:20 UT.

A spectrogram of electric field intensities obtained from the Galileo plasma instrument in the vicinity of Io is shown in Fig. 4. The red color in the spectrum represents strongest emission while the blue color represents the weakest emission. The range from dark blue to bright red is 70 dB. The time range is chosen from 01:08:20 UT to 01:38:20 UT so that the auroral hiss-like emission can be shown clearly. This radiation occurs from 01:16:00 UT to 01:20:00 UT and spans a frequency range from about 1 kHz to 40 kHz. The radiation has an asymmetrical funnel-shaped low frequency cutoff that decreases monotonically from about 40 kHz at 01:16:00 UT to 10 kHz at 01:20:00 UT. The electron cyclotron frequency shown by the white line marked $f_{ce}$ was computed from on-board magnetic field measurements using the equation $f_{ce} = 28\,B\,Hz$ where B is in nT. It has a value about 58 kHz and the proton cyclotron frequency $f_{ci}$ has a value of about 32 Hz. The electron plasma frequency, shown by the white line marked $f_{pe}$, is also shown in the figure and has a value about 600 kHz during the period of interest. As can be seen the following inequalities exist among the proton cyclotron frequency, the observed emission frequency $f$, the electron cyclotron frequency, and the electron plasma frequency:

$$f_{ci} \ll f \ll f_{ce} < f_{pe} \,. \tag{1}$$

For these parameters the only possible mode of propagation in the frequency range of interest is the whistler mode. As will be shown shortly these inequalities will allow us to greatly simplify the cold plasma dispersion relation which will be used later to perform ray-path calculations. The plots of X, Y, Z components of the magnetic field in nT for the same time range in Fig. 4, is shown in Fig. 5. From these plots it can be seen that all three components of the magnetic field are smooth and slowly varying except in the interval between 01:15:00 UT to 01:32:00 UT which corresponds to the time range when the spacecraft was in the vicinity of Io. In this interval saw-shaped perturbations are obviously observed in the $B_x$ and $B_y$ plots with perturbations amplitudes of about $\Delta B_x = 600\,nT$ (in a background of -300 nT) and $\Delta B_y = 300\,nT$ (relative to a background of -300 nT). Obvious abrupt changes in magnetic field occurred at about 01:21:00 UT to 01:29:00 UT. According to Ampere's law $\nabla \times \boldsymbol{B} = \mu_0 \boldsymbol{J}$, those changes indicate that the spacecraft crossed two intense current sheets, one near the inner boundary of Io, and the other near the outer boundary of Io. In the region between the two major current sheets, the z-component of the magnetic field increased (decreased in magnitude) gradually with small oscillations. After the second current sheet crossing, the $B_z$ field drops down to an equilibrium value of about -1650 nT which is slightly larger than the field (-1900 nT) that was present during the approach to Io. Comparing with the electric field spectrum, it can be seen that the auroral hiss-liked emission occurred when the magnetic perturbation started. The vertex of the funnel-shaped emission occurred almost exactly at the same moment as the first major magnetic field discontinuity.

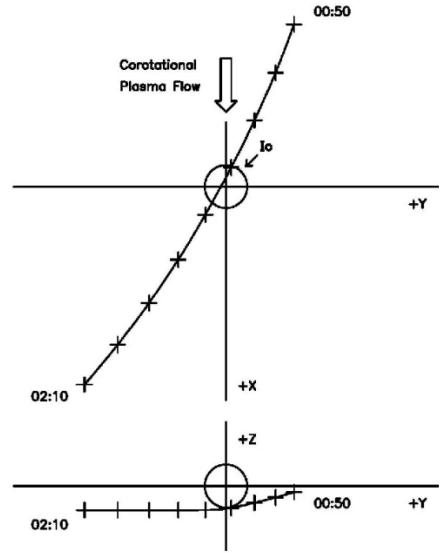

Figure 3. Trajectory of Galileo during the Io flyby in Oct. 16[th], 2001.

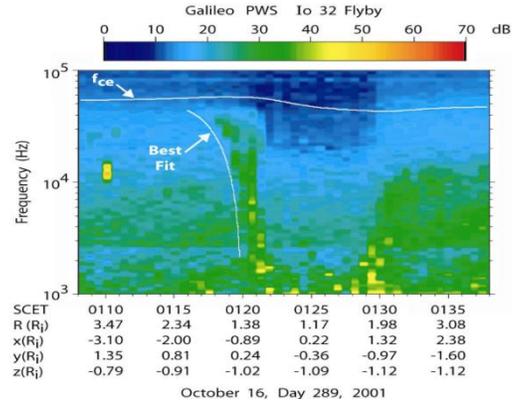

Figure 4. Time-Frequency spectrum of the electric field for the time series from 01:08:20 UT to 01:38:20 UT, Oct. 16[th], 2001.

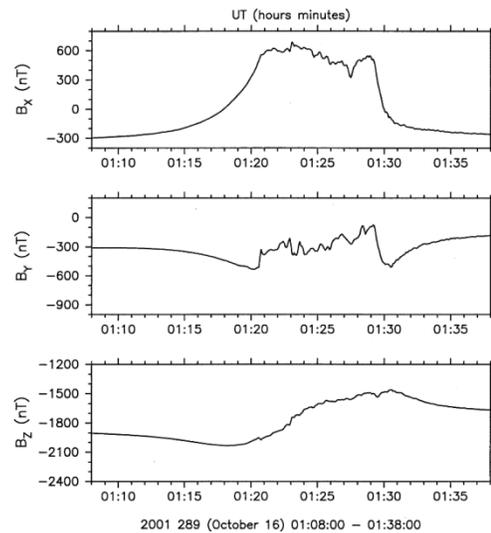

Figure 5. Magnetic field components versus spacecraft event time, Oct. 16[th], 2001.

These facts can be explained as follows: when the spacecraft approached the current sheet auroral hiss-liked radiation generated by the current was first detected; the radiation was continuously received with strongest wide spectrum radiation occurring as the spacecraft was right in the current sheet. These data suggest that the auroral hiss is closely associated with the current that cause discontinuity in the magnetic field. This fact gives further evidence of the presence of a field aligned current flow connecting Io with Jupiter as proposed by Goldreich and Lynden-Bell (1968). Similar magnetic perturbations of about 5% were also detected earlier by Voyager 1 when the spacecraft crossed Io's magnetic flux tube about 11 $R_{Io}$ below Io.

**B. The Unipolar Inductor Model**

There is a motional induced charge separation in a conductor moving across magnetic field lines. This charge may be conducted away, resulting in a dc current flow through the conductor if it moves through plasma. The generation of Alfven waves is a mechanism particularly effective for circulating the charge for very large conductors moving in or above the planetary ionosphere.

Alfven waves were first postulated by Alfven using the theory of magnetohydrodynamics or MHD [1]. MHD refers to a highly conducting fluid with the presence of a magnetic field. In the case of an incompressible MHD fluid, Alfven found that there is one wave mode that can propagate. This wave often refers to as a shear Alfven wave, is primarily electromagnetic in nature. The wave is characterized by a perturbation magnetic field which is transverse to the undisturbed magnetic field and propagates along the direction of undisturbed field. Also the fluid is perturbed in a direction transverse to the undisturbed magnetic field. Often the shear Alfven wave is described as being analogous to a wave on a taut string where the Maxwell stress provides the tension. Herlofson and Van de Hulst first demonstrated that if an MHD fluid is allowed to be compressible two additional wave modes are possible: the fast and slow magnetosonic modes [2]. These two modes have both acoustic and electromagnetic character and are also sometimes referred to as Alfven waves. This section is primarily concerned with shear Alfven waves which will be referred to simply as Alfven waves and the other MHD modes will be referred to as magnetosonic waves [2,3,4]. The goal of this section is to review the theory of Alfven wave generation by a moving object in a magnetized plasma environment and the relevant current sheet due to this motion.

As a basic the model of Drell [1965] for a satellite is presented. Drell represent the satellite as a perfect conductor and assume the satellite moves through plasma with a uniform magnetic field. The approximation of cold plasma is made; and linear perturbation theory is applied. Further approximations appropriate to obtain MHD results are made. Then a wave equation for an Alfven wave is obtained.

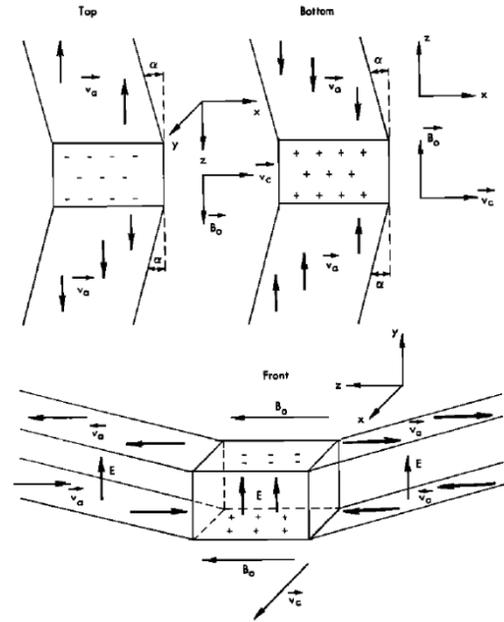

**Figure 6.** An ideal conductor moving through plasma in a direction perpendicular to the magnetic field. Charge separation occurs in the conductor from Ref. 11.

Consider a perfect conductor moving in a collisionless plasma in a direction perpendicular to a uniform magnetic field $\boldsymbol{B}_0$. As shown in Fig. 6, the magnetic field $\boldsymbol{B}_0$ is in the z direction, and the velocity $\boldsymbol{V}_c$ is in the x direction. In the conductor a motional electric field, $\boldsymbol{E} = \boldsymbol{V}_c \times \boldsymbol{B}_0$ will be just canceled by the charge separation shown in Fig. 6. Also assume that the conductor has no work function to prevent the flow of electrons from the conductor's surface into the plasma. In addition, it must be considered that for collisionless plasma the conductivity parallel to magnetic field line is large, and the conductivity perpendicular to it is small ($\frac{\sigma_\parallel}{\sigma_\perp} \gg 1$).

It can be shown that a single particle immersed in a constant, uniform equilibrium magnetic field $\boldsymbol{B}_0 = B_0 \hat{z}$ and subject to a small-amplitude wave with electric field $\sim exp(i\boldsymbol{k}.\boldsymbol{x} - i\omega t)$ has the velocity:

$$\widetilde{\boldsymbol{V}}_\sigma = \frac{iq_\sigma}{\omega m_\sigma} \left[ \widetilde{E}_z \hat{z} + \frac{\widetilde{\boldsymbol{E}}_\perp}{1 - \omega_{c\sigma}^2/\omega^2} - \frac{i\omega_{c\sigma}}{\omega} \frac{\hat{z} \times \widetilde{\boldsymbol{E}}}{1 - \omega_{c\sigma}^2/\omega^2} \right] e^{i\boldsymbol{k}.\boldsymbol{x} - i\omega t} \,. \qquad (2)$$

The tilde ~ denotes a small-amplitude oscillatory quantity with space time dependence $exp(i\mathbf{k}.\mathbf{x} - i\omega t)$; this phase factor may or may not be explicitly written, but should always be understood to exist for a tilde-denoted quantity. The three terms in Eq. (2) are respectively:

1) *The parallel quiver velocity*: this quiver velocity is the same as the quiver velocity of an unmagnetized particle, but is restricted to parallel motion. Because the magnetic force $q(\mathbf{V} \times \mathbf{B})$ vanishes for motion along the magnetic field, motion parallel to $\mathbf{B}$ in a magnetized plasma is identical to motion in an unmagnetized plasma.

2) *The generalized polarization drift*: this motion has a resonance at the cyclotron frequency but at low frequencies such that $\omega \ll \omega_{c\sigma}$, it reduces to the polarization drift $\mathbf{V}_{p\sigma} = m_\sigma \dot{\mathbf{E}}_\perp / q_\sigma B^2$.

3) *The generalized $\mathbf{E} \times \mathbf{B}$ drift*: this also has a resonance at the cyclotron frequency and for $\omega \ll \omega_{c\sigma}$ reduces to the drift $\mathbf{V}_E = \frac{\mathbf{E} \times \mathbf{B}}{B^2}$.

The particle velocities given in Eq. (2) produce a plasma current density:

$$\tilde{\mathbf{J}} = \sum_\sigma n_{0\sigma} q_\sigma \tilde{\mathbf{V}}_\sigma = i\varepsilon_0 \sum_\sigma \frac{\omega_{p\sigma}^2}{\omega} \left[ \tilde{E}_z \hat{z} + \frac{\tilde{\mathbf{E}}_\perp}{1 - \omega_{c\sigma}^2/\omega^2} - \frac{i\omega_{c\sigma}}{\omega} \frac{\hat{z} \times \tilde{\mathbf{E}}}{1 - \omega_{c\sigma}^2/\omega^2} \right] e^{i\mathbf{k}.\mathbf{x} - i\omega t} . \qquad (3)$$

where $\omega_{p\sigma} = \frac{n_{0\sigma} q_\sigma^2}{\varepsilon_0 m_\sigma}$. If these plasma currents are written out explicitly, then Ampere's law has the form:

$$\nabla \times \tilde{\mathbf{B}} = \mu_0 \tilde{\mathbf{J}} + \mu_0 \varepsilon_0 \frac{\partial \tilde{\mathbf{E}}}{\partial t} = \mu_0 \left( i\varepsilon_0 \sum_\sigma \frac{\omega_{p\sigma}^2}{\omega} \left[ \tilde{E}_z \hat{z} + \frac{\tilde{\mathbf{E}}_\perp}{1 - \omega_{c\sigma}^2/\omega^2} - \frac{i\omega_{c\sigma}}{\omega} \frac{\hat{z} \times \tilde{\mathbf{E}}}{1 - \omega_{c\sigma}^2/\omega^2} \right] - i\omega \varepsilon_0 \tilde{\mathbf{E}} \right) . \qquad (4)$$

where a factor $exp(i\mathbf{k}.\mathbf{x} - i\omega t)$ is implicit.

The cold plasma wave equation is established by combining Ampere's and Faraday's law in a manner similar to the method used for vacuum electromagnetic waves. However, before doing so, it is useful to define the dielectric tensor $\vec{K}$. This tensor contains the information in the right hand side of Eq. (4) so that this equation is written as:

$$\nabla \times \mathbf{B} = \mu_0 \varepsilon_0 \frac{\partial}{\partial t} \left( \vec{K}.\mathbf{E} \right) . \qquad (5)$$

Where $\vec{K}.\mathbf{E}$ is:

$$\begin{aligned} \vec{K}.\mathbf{E} &= \mathbf{E} - \sum_{\sigma=i,e} \frac{\omega_{p\sigma}^2}{\omega^2} \left[ \tilde{E}_z \hat{z} + \frac{\tilde{\mathbf{E}}_\perp}{1 - \omega_{c\sigma}^2/\omega^2} - \frac{i\omega_{c\sigma}}{\omega} \frac{\hat{z} \times \tilde{\mathbf{E}}}{1 - \omega_{c\sigma}^2/\omega^2} \right] \\ &= \begin{bmatrix} S & -iD & 0 \\ iD & S & 0 \\ 0 & 0 & P \end{bmatrix} . \mathbf{E} . \end{aligned} \qquad (6)$$

And the elements of dielectric tensor are:

$$S = 1 - \sum_{\sigma=i,e} \frac{\omega_{p\sigma}^2}{\omega^2 - \omega_{c\sigma}^2}, \quad D = \sum_{\sigma=i,e} \frac{\omega_{c\sigma}}{\omega} \frac{\omega_{p\sigma}^2}{\omega^2 - \omega_{c\sigma}^2}, \quad P = 1 - \sum_{\sigma=i,e} \frac{\omega_{p\sigma}^2}{\omega^2} . \qquad (7)$$

The terms S and D can be decomposed into a sum and a difference using the relations:

$$S = \frac{R+L}{2}, D = \frac{R-L}{2}. \quad (8)$$

where R and L are defined by

$$R = 1 - \sum_{\sigma=i,e} \frac{\omega_{p\sigma}^2}{\omega(\omega + \omega_{c\sigma})}, L = 1 - \sum_{\sigma=i,e} \frac{\omega_{p\sigma}^2}{\omega(\omega - \omega_{c\sigma})}. \quad (9)$$

So that R diverges when $\omega = -\omega_{c\sigma}$ whereas L diverges when $\omega = \omega_{c\sigma}$. Since $\omega_{c\sigma} = \frac{q_\sigma B}{m_\sigma}$, the ion cyclotron frequency is positive and the electron cyclotron frequency is negative. Hence, R diverges at the electron cyclotron frequency, whereas L diverges at the ion cyclotron frequency. When $\omega \to \infty$, both $R, L \to 1$. In the limit $\omega \to 0$, evaluation of R, L must be done very carefully, since

$$\frac{\omega_{p\sigma}^2}{\omega_{c\sigma}} = \frac{n_\sigma q_\sigma^2}{\varepsilon_0 m_\sigma} \frac{m_\sigma}{q_\sigma B} = \frac{n_\sigma q_\sigma}{\varepsilon_0 B}, \quad (10)$$

so that

$$\frac{\omega_{pi}^2}{\omega_{ci}} = -\frac{\omega_{pe}^2}{\omega_{ce}}. \quad (11)$$

thus

$$\lim_{\omega \to 0} R, L = 1 - \frac{1}{\omega}\left[\frac{\omega_{pi}^2}{(\omega \pm \omega_{ci})} + \frac{\omega_{pe}^2}{(\omega \pm \omega_{ce})}\right]$$
$$= 1 - \frac{\omega_{pi}^2 + \omega_{pe}^2}{\omega_{ci}\omega_{ce}}$$
$$\cong 1 - \frac{n_e q_e^2}{\varepsilon_0 m_e}\frac{m_i}{q_i B}\frac{m_e}{q_e B} \quad (12)$$
$$= 1 + \frac{\omega_{pi}^2}{\omega_{ci}^2}$$
$$= 1 + \frac{c^2}{V_A^2}.$$

These simplifications imply that to obtain we are concerned with oscillations involving bulk motion of the plasma. Then the time scale for fluid motion should be longer than the ion cyclotron gyration's time or equivalently $\omega \ll \omega_{ci}$. Long time scales are required, because for time scales shorter than the ion cyclotron gyration's time, the electron and ion behave quite differently and the approximation for single fluid motion is not appropriate. By using Eq. (6) and (12) the dielectric tensor has the following form:

$$\overleftrightarrow{K} = \begin{bmatrix} 1 + \dfrac{c^2}{V_A^2} & 0 & 0 \\ 0 & 1 + \dfrac{c^2}{V_A^2} & 0 \\ 0 & 0 & 1 - \sum_{\sigma=i,e} \dfrac{\omega_{p\sigma}^2}{\omega^2} \end{bmatrix}. \tag{13}$$

The relation between the conductivity and the dielectric tensor is $\overleftrightarrow{K} = \overleftrightarrow{I} - \dfrac{\overleftrightarrow{\sigma}}{i\varepsilon_0\omega}$. This relation was obtained by writing the Ampere's law in microscopic and macroscopic forms. So that the conductivity tensor is:

$$\overleftrightarrow{\sigma} = \begin{bmatrix} i\varepsilon_0\omega \dfrac{c^2}{V_A^2} & 0 & 0 \\ 0 & i\varepsilon_0\omega \dfrac{c^2}{V_A^2} & 0 \\ 0 & 0 & -i\varepsilon_0\omega \sum_{\sigma=i,e} \dfrac{\omega_{p\sigma}^2}{\omega^2} \end{bmatrix} \text{ or } \overleftrightarrow{\sigma} = \begin{bmatrix} \sigma_\perp & 0 & 0 \\ 0 & \sigma_\perp & 0 \\ 0 & 0 & \sigma_\parallel \end{bmatrix}. \tag{14}$$

The requirement that $\dfrac{\sigma_\parallel}{\sigma_\perp} \gg 1$ then becomes $\omega^2 \ll \dfrac{\omega_{p\sigma}^2 V_A^2}{c^2}$ which is consistent with the low frequency assumption made. In order to obtain equations for MHD waves, we next consider Maxwell's equations with the expression for $\overleftrightarrow{K}$ from Eq. (13). Also the field strengths are linearized by setting $\boldsymbol{B} = \boldsymbol{B}_0 + \boldsymbol{B}_1$, where $\boldsymbol{B}_0$ is the undisturbed uniform magnetic field and setting $\boldsymbol{E} = \boldsymbol{E}_1$. Source terms $\rho(\boldsymbol{r},t)$ and $\boldsymbol{J}(\boldsymbol{r},t)$, external to the plasma are assumed. These charge and current source are provided by moving conductor. For the present, $\rho(\boldsymbol{r},t)$ is left completely general; but $\boldsymbol{J}(\boldsymbol{r},t) = \boldsymbol{J}(x - V_c t, t)$. That is, the source current is the current through the moving conductor, and thus the current must depend on $(x - V_c t)$. The dielectric tensor of plasma is represented by:

$$\begin{aligned} \boldsymbol{D} &= \overleftrightarrow{\varepsilon} \boldsymbol{E} \\ \boldsymbol{D} &= \varepsilon_0 \overleftrightarrow{K} \boldsymbol{E} \\ \overleftrightarrow{\varepsilon} &= \varepsilon_0 \overleftrightarrow{K} \end{aligned} \tag{15}$$

By some simplification the $\varepsilon_\perp, \varepsilon_\parallel$ are as below.

$$\begin{aligned} \varepsilon_\perp &= \varepsilon_0 \left(1 + \dfrac{c^2}{V_A^2}\right) \\ \varepsilon_\parallel &= \varepsilon_0 \left(1 - \sum_{\sigma=i,e} \dfrac{\omega_{p\sigma}^2}{\omega^2}\right) \end{aligned} \tag{16}$$

Maxwell's equations then, in Fourier transformed, are the following:

$$\begin{aligned} -\boldsymbol{k}(\boldsymbol{k}.\boldsymbol{E}_1) + k^2 \boldsymbol{E}_1 &= i\omega\mu_0 \boldsymbol{J}(\boldsymbol{k})\delta(\omega - \boldsymbol{k}.\boldsymbol{V}_c) + \mu_0 \omega^2 \varepsilon_0 \overleftrightarrow{K} \boldsymbol{E}_1 \\ \varepsilon_\parallel k_\parallel E_\parallel + \varepsilon_\perp \boldsymbol{k}_\perp . \boldsymbol{E}_\perp &= \dfrac{\rho}{i} \rightarrow \varepsilon_\parallel k_\parallel E_\parallel + \varepsilon_\perp k_\perp E_\perp^l = \dfrac{\rho}{i} \end{aligned} \tag{17}$$

In above equations $\boldsymbol{E}_\perp = \boldsymbol{E}_\perp^l + \boldsymbol{E}_\perp^{tr}$, where $\boldsymbol{k}_\perp.\boldsymbol{E}_\perp = k_\perp E_\perp^l$ and $\boldsymbol{k}_\perp \times \boldsymbol{E}_\perp = \boldsymbol{k}_\perp \times \boldsymbol{E}_\perp^{tr}$. By using Eqs. (16) and (17) the following equations are obtained:

$$\left[\omega^2 \mu_0 \varepsilon_\perp - k_\parallel^2 - k_\perp^2 \frac{\varepsilon_\perp}{\varepsilon_\parallel}\right] E_\perp^l = \frac{\omega \mu_0}{i} \frac{\boldsymbol{J}(\boldsymbol{k}).\boldsymbol{k}_\perp}{k_\perp} \delta(\omega - \boldsymbol{k}.\boldsymbol{V}_c) - \frac{k_\perp \rho}{i\varepsilon_\parallel} , \qquad (18)$$

and

$$[\omega^2 \mu_0 \varepsilon_\perp - k_\parallel^2 - k_\perp^2] E_\perp^{tr} = \frac{\omega \mu_0}{i} \frac{|\boldsymbol{J}(\boldsymbol{k}) \times \boldsymbol{k}_\perp|}{k_\perp} \delta(\omega - \boldsymbol{k}.\boldsymbol{V}_c) . \qquad (19)$$

Eq. (18) represents the Alfven wave. For $\frac{\varepsilon_\perp}{\varepsilon_\parallel} \ll 1$, the equation reduces to a one dimensional equation for a transverse electric field propagating along the $\boldsymbol{B}_0$ direction with velocity about $V_A$ and frequency $\omega = \boldsymbol{k}.\boldsymbol{V}_c$. The wavelength parallel to $V_c$ is about L, the length of the conductor along the direction of motion, so that $\omega \cong \frac{2\pi V_c}{L}$. Eq. (19) represents a magnetosonic wave, also with velocity $V_A$. But this wave propagates isotropically, and thus the energy falls away as $\frac{1}{r^2}$. In contrast, the Alfven wave energy is concentrated along one direction. So the Alfven wave is more important.

Now Eq. (18) is transformed back to real space:

$$\left[-\frac{1}{V_A^2} \frac{\partial^2}{\partial t^2} + \frac{\partial^2}{\partial z^2}\right] \nabla_\perp . \boldsymbol{E} = \mu_0 \frac{\partial}{\partial t} \nabla_\perp . \boldsymbol{J}(x - V_c t, y, z) . \qquad (20)$$

It follows from this form of the equation that the wave field $\boldsymbol{E}_1$ is a function of variables $y, x - V_c t \pm \left(\frac{V_c}{V_A}\right) z$ and therefore $\boldsymbol{B}$ is a function of the same variables. Thus, as stated earlier, the perturbation fields occur in Alfven wings which are at an angle $\alpha = tan^{-1} \frac{V_c}{V_A}$ to the direction of $\boldsymbol{B}_0$. Since the tangential component of the electric field $\boldsymbol{E}$ is continuous across the surface of the conductor, the electric field between the wings has the same magnitude as that in the conductor: $-(\boldsymbol{V}_c \times \boldsymbol{B}_0)$. From the Maxwell's perturbed equations is can be determined that the magnitude of the perturbation magnetic field is $\boldsymbol{B}_1 = \underbrace{\left(\frac{V_c}{V_A}\right)}_{M_A} \boldsymbol{B}_0$. Thus, the linear theory is valid when Alfven Mach number $M_A \ll 1$. The plasma in the wings will experience an $\boldsymbol{E} \times \boldsymbol{B}$ drift that will approach the velocity of the conductor $V_c$. That is the reason, the plasma between the wings will move with the conductor. Similar process occurs in the Jovian system; due to the Io's motion in the plasma torus of the Jupiter and in the magnetosphere of this giant solar system planet the Alfven waves and the corresponding current sheets are produced and phenomena will be used to analyze sheet source aligned along the magnetic field lines.

Now we proceed to find an expression for group velocity direction. The definition of dielectric tensor means that Maxwell's equations, the Lorentz equation, and the plasma currents can now be summarized in just two coupled equations, namely

$$\begin{aligned} \nabla \times \boldsymbol{B} &= \frac{1}{c^2} \frac{\partial}{\partial t} (\overleftrightarrow{K}.\boldsymbol{E}) \\ \nabla \times \boldsymbol{E} &= -\frac{\partial \boldsymbol{B}}{\partial t} . \end{aligned} \qquad (21)$$

The cold plasma wave equation is obtained by taking the curl of $\nabla \times \boldsymbol{E} = -\frac{\partial \boldsymbol{B}}{\partial t}$ and then substituting for $\nabla \times \boldsymbol{B}$ using $\nabla \times \boldsymbol{B} = \frac{1}{c^2} \frac{\partial}{\partial t}(\overleftrightarrow{K}.\boldsymbol{E})$ to obtain:

$$\nabla \times (\nabla \times \boldsymbol{E}) = -\frac{1}{c^2} \frac{\partial^2}{\partial t^2} (\overleftrightarrow{K}.\boldsymbol{E}) . \qquad (22)$$

Since a phase dependence $exp(i\mathbf{k}.\mathbf{x} - i\omega t)$ is assumed, this can be written in algebraic form as

$$\mathbf{k} \times (\mathbf{k} \times \mathbf{E}) = -\frac{\omega^2}{c^2}\vec{K}.\mathbf{E} \ . \tag{23}$$

By using $\mathbf{n} = \left(\frac{c\mathbf{k}}{\omega}\right)$ Eq. (10) becomes

$$\mathbf{nn}.\mathbf{E} - n^2\mathbf{E} + \vec{K}.\mathbf{E} = 0 , \tag{24}$$

which is essentially a set of three homogenous equations in the three components of $\mathbf{E}$. The refractive index can be decomposed into parallel and perpendicular components relative to the equilibrium magnetic field $\mathbf{B}_0 = B_0\hat{z}$. For convenience the $x$ axis of the coordinate system is defined to lie along the perpendicular component of $\mathbf{n}$ so that $n_y = 0$ by assumption. This simplification is possible for a spatially uniform equilibrium only; if the plasma is non-uniform in the $x - y$ plane, there can be a real distinction between $x$ and $y$ direction propagation and the refractive index in the $y$ direction cannot be simply defined away by choice of coordinate system.

To set the stage for obtaining a dispersion relation, Eq. (11) is written in matrix form as

$$\begin{bmatrix} S - n_z^2 & -iD & n_x n_z \\ iD & S - n^2 & 0 \\ n_x n_z & 0 & P - n_x^2 \end{bmatrix} . \begin{bmatrix} E_x \\ E_y \\ E_z \end{bmatrix} = 0 . \tag{25}$$

It is useful to introduce a spherical coordinate system in $k$ space with $\hat{z}$ defining the axis and $\theta$ the polar angle. Thus, the Cartesian components of the refractive index are related to the spherical components by

$$n_x = nsin\theta, n_z = ncos\theta, n^2 = n_x^2 + n_y^2 \ . \tag{26}$$

And so Eq. (12) becomes

$$\begin{bmatrix} S - n^2 cos^2\theta & -iD & n^2 sin\theta cos\theta \\ iD & S - n^2 & 0 \\ n^2 sin\theta cos\theta & 0 & P - n^2 sin^2\theta \end{bmatrix} . \begin{bmatrix} E_x \\ E_y \\ E_z \end{bmatrix} = 0 . \tag{27}$$

This equation has non-trivial solution if and only if the determinant of the matrix is zero. After some algebra this determinant can be written as:

$$An^4 - Bn^2 + C = 0 . \tag{28}$$

where

$$\begin{aligned} A &= Ssin^2\theta + Pcos^2\theta , \\ B &= (S^2 - D^2)sin^2\theta + PS(1 + cos^2\theta) , \\ C &= P(S^2 - D^2) = PRL \ . \end{aligned} \tag{29}$$

By sorting out in the dispersion relation, Eq. (28), the following equation is obtained:

$$tan^2\theta = \frac{-P(n^2 - R)(n^2 - L)}{(Sn^2 - RL)(n^2 - P)}. \tag{30}$$

To obtain a relation for the index of refraction of the whistler mode: first, we ignore the ions terms since they response much more slowly than the electron in the frequency of interest due to their heavy masses. Second, we assume that both the wave frequency and the electron cyclotron frequency are both much less than the electron plasma frequency ($\omega^2 \ll \omega_p^2, \omega_c^2 \ll \omega_p^2$). These conditions are satisfied in our case, because the auroral hiss emission region is well below the electron cyclotron frequency ($f < f_c$), as shown in Fig. 4, and the plasma frequency is about ten times greater than cyclotron frequency ($f_p \approx 590 KHz \gg f_c \approx 58 KHz$). With this assumption and the corrections applied to expressions for R, L, D, S, and P, it is easy to obtain a relation for the index of refraction as the following:

$$n^2 = \frac{\omega_p^2}{\omega(\omega_c \cos\theta - \omega)}. \tag{31}$$

Three polar plots of this equation for frequency $f_1, f_2,$ and $f_3$ are shown in Fig. 1. The resonance cone is defined as the locus of points where the index of refraction goes to infinity. The corresponding wave normal angle is called the resonance cone angle, which in this case is given by $cos\theta_{res} = \frac{\omega}{\omega_c}$. Since the group velocity of wave propagation is perpendicular to the index of refraction surface, then the angle $\psi$, between the direction of the group velocity and the magnetic field is given by $\psi = 90^o - \theta_{res}$. It follows then that the group velocity direction, which is the direction of energy flow is given by:

$$sin\psi = \frac{f}{f_c}. \tag{32}$$

It is obvious that the higher frequencies have larger angles of propagation. Fig. 7 shows a simple point source near Io. When an observer approaches the radiation source from left, the highest radiation frequency $f_3$ is received; then the smaller frequencies $f_2$, and $f_1$ will be detected respectively.

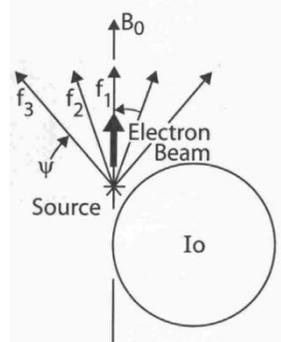

**Figure 7. Propagation of whistle mode waves from a point source.**

In order to explain the funnel-shaped cutoff characteristic of the auroral hiss, we consider a simple two dimensional model. Two assumptions are made to simplify the model; first that the magnetic field is uniform everywhere and is perpendicular to the trajectory of the spacecraft, and the second, that the radiation source is a point source. To carry out a simplified analysis, we introduced coordinate $x$, which is a perpendicular distance from the spacecraft to the magnetic field line through the source and the distance $h$, which is the height of the spacecraft above the source as shown in Fig. 8.

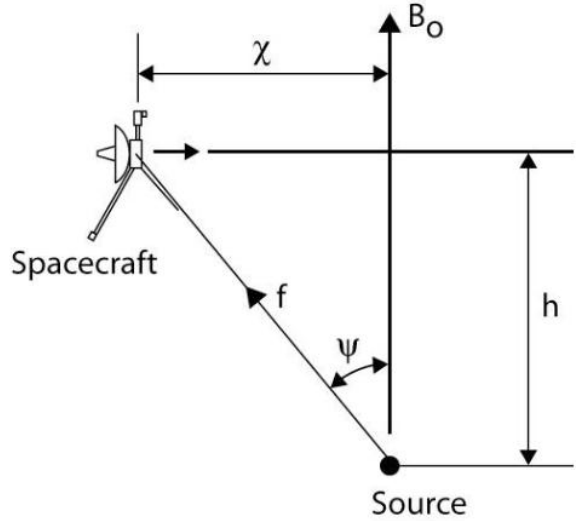

**Figure 8. The geometry of the produce of a funnel-shaped low frequency cutoff.**

Then a simple geometry shows:

$$tan\psi = \frac{x}{h}. \tag{33}$$

Also for Eq. (32) it is possible to obtain:

$$tan\psi = \frac{f}{\sqrt{f_c^2 - f^2}}. \tag{34}$$

Equating Eqs. (33) and (34) gives the low frequency cutoff as the following:

$$f^2 = f_c^2 - \frac{h^2}{x^2 + h^2} f_c^2. \tag{35}$$

This equation is a hyperbola with an upper frequency limit $f_c$.

### C. Calculation of point source model and the sheet source model

Under the assumptions $\omega^2 \ll \omega_p^2$, $\omega_c^2 \ll \omega_p^2$ we have $sin\psi = \frac{f}{f_c}$, where $\psi$ is the angle between the limiting ray path direction and the magnetic field. Fig. 9 shows the geometric relation to locate the point source of emission. $(x, y, z)$ is an arbitrary point on the trajectory of the Galileo; $(x_s, y_s, z_s)$ represents the position of the emission source; $(x_0, y_0, z_0)$ is a point when the spacecraft is on the magnetic field line through the source; or is a point where the lowest frequency of the emission of the cutoff boundary is received. The low frequency apex of the emission occurs at about 01:20:00 UT (as shown in Fig. 4), which corresponds to spacecraft coordinates $(x_0, y_0, z_0) = (-0.893, 0.24, -1.019)$. All of the values is a ration of the Io's radius $R_{Io}$. The height of the source $h = |\boldsymbol{R}_0 - \boldsymbol{R}_s|$ is adjusted until a best-fit cutoff boundary is found. If the height of the emission source $h$ is given, then the coordinates of the source $(x_s, y_s, z_s)$ can be calculated by simple geometry relations:

$$\frac{x_0 - x_s}{B_x} = \frac{y_0 - y_s}{B_y} = \frac{z_0 - z_s}{B_z} = \frac{h}{B} \Rightarrow \begin{cases} x_s = x_0 - \frac{h}{B}B_x \\ y_s = y_0 - \frac{h}{B}B_y \\ z_s = z_0 - \frac{h}{B}B_z \end{cases}. \quad (36)$$

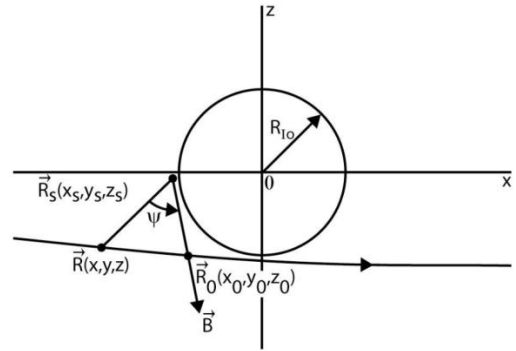

**Figure 9. The geometric relation for locating the point source position when using best-fit propagation cutoff.**

The required data for magnetic field obtained from Fig. 5. For each point in the range 01:15:00 UT to 01:20:00 UT on the trajectory $R(x, y, z)$ the angle $\psi$ between $\boldsymbol{R} - \boldsymbol{R}_s$ and $\boldsymbol{R}_0 - \boldsymbol{R}_s$ should be calculated by using the following relation:

$$cos\psi = \frac{(\boldsymbol{R} - \boldsymbol{R}_s).(\boldsymbol{R}_0 - \boldsymbol{R}_s)}{|\boldsymbol{R} - \boldsymbol{R}_s||\boldsymbol{R}_0 - \boldsymbol{R}_s|} = \frac{(x - x_s)(x_0 - x_s) + (y - y_s)(y_0 - y_s) + (z - z_s)(z_0 - z_s)}{\sqrt{(x - x_s)^2 + (y - y_s)^2 + (z - z_s)^2}\sqrt{(x_0 - x_s)^2 + (y_0 - y_s)^2 + (z_0 - z_s)^2}} \quad (37)$$

Using the above equation we get the time dependent of the cutoff frequency ($f(t) = f_c sin\psi = f_c\sqrt{1 - cos^2\psi}$). Fig. 10 shows the results for different value of $h$.

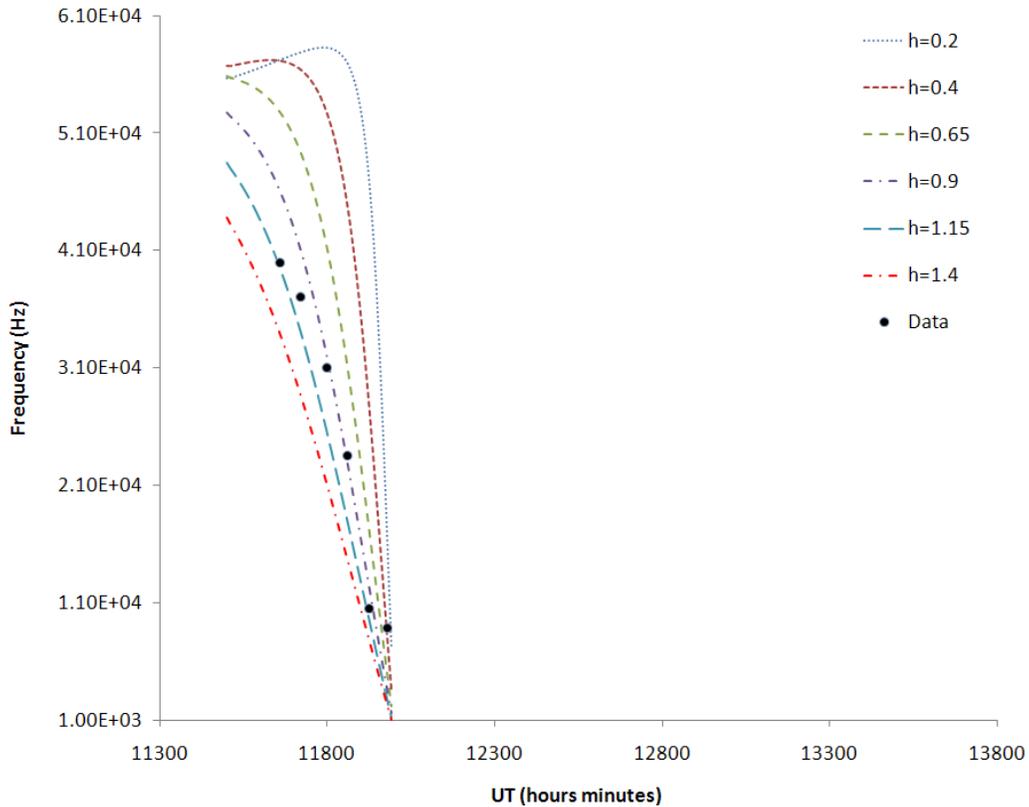

**Figure 10.** Cutoff Frequencies vs. universal times for different height between the Galileo and the point source.

The solid points are sampled from Fig. 4 from the cutoff boundary of the spectrum. The wide range of $h$ indicates that the source does not have a sharp defined low altitude boundary. That is reasonable since the emission spectrum in Fig. 4 does not have a sharply defined frequency time boundary. The best fit to the cutoff frequency data gives $h = 0.9$, which corresponds to the source position at coordinates $(x_s, y_s, z_s) = (-1.0434, 0.469, -0.16)$.

The small value of $z_s$ indicates that the source lies near the equator of Io; in the region where Jupiter's magnetic field is tangent to the surface if Io. The current system between the Jupiter and the Io which is probably the reason of the auroral hiss is shown in Fig. 11. The electromotive force which is across the Io's radial diameter drives a current that flows on the surface of the magnetic flux tube connecting Io with Jupiter at geographical colatitude for the northern foot of $\theta_i = 24°$.

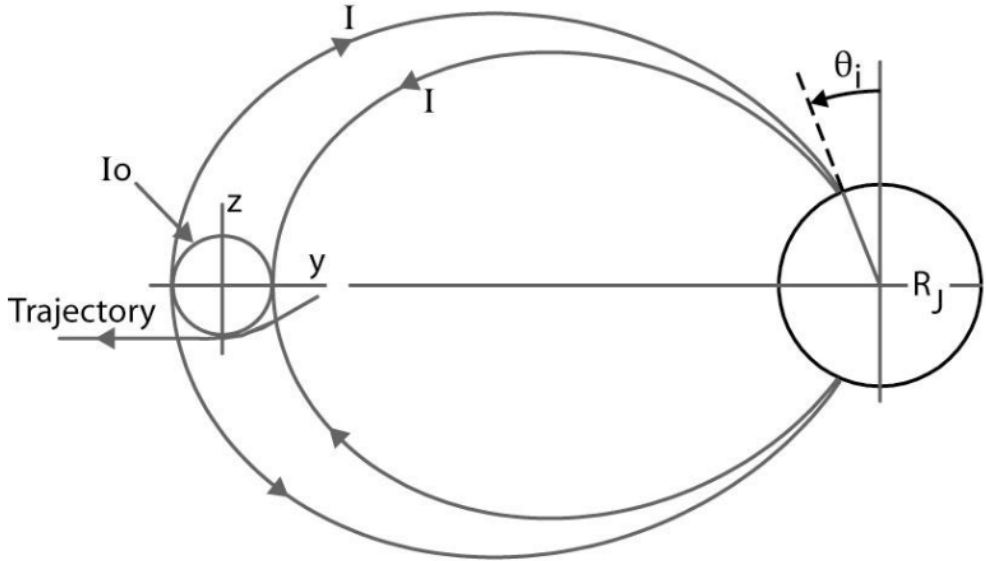

**Figure 11. The current circuit of Io and Jupiter and the trajectory of Galileo are shown.**

Since the frequency-time spectrum of the radiation is filled in instead of being a sharp line, it is likely that the source of the emission is either a line or a sheet source. Applying the unipolar inductor model, we consider the possibility that the source is a cylindrical current sheet (Fig. 12 and 13).

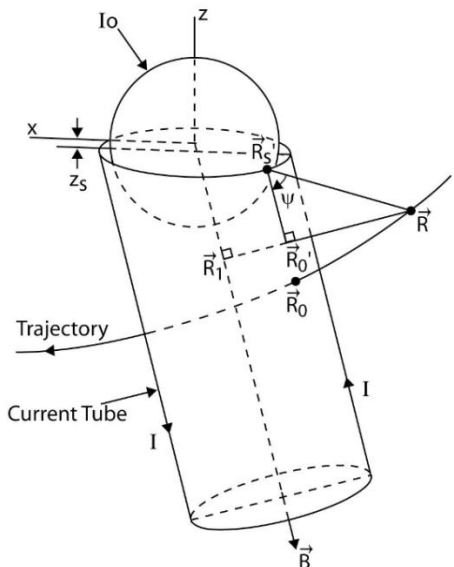 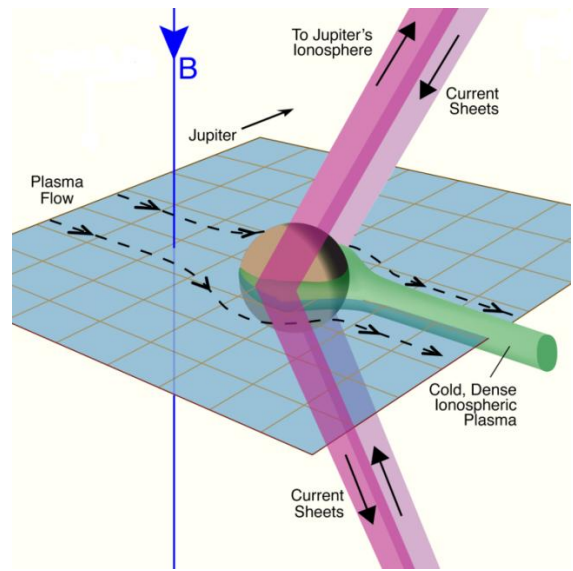

**Figure 12. Sheet emission source model.**     **Figure 13. 3-D model of Io's current sheet.**

The axis of the cylinder is in the direction of the magnetic field line at $R_0$ through the center of the Io with radius $r = \sqrt{x_s^2 + y_s^2 + z_s^2} = 1.15 > 1$. This issue is reasonable since the current sheet is most likely produced in the ionosphere of the Io. Because we have no information of the magnetic field out of the trajectory, we simply assume that the local current sources responsible for the radiation are also along $B_0$ at $R_0$. Also for simplicity, plane P has drawn with its normal along z direction through point $R_s$. The intersection between the plane P and the cylindrical current sheet makes a curve C. For every point on the trajectory, the normal of the current cylinder $R_0'R$ has drawn. Note $R_0'$ lies on the surface of the cylinder. Line $R_s'R_0'$ is parallel to $B_0$ and intersects with curve C at point $R_s'$. The coordinates of $R_0' = (x_0', y_0', z_0')$ and $R_s' = (x_s', y_s', z_s')$ can be calculated as follows: Define the unit vector of $B_0$ as $n_B = (a, b, c) = (0.167, -0.255, -0.952)$, then the coordinates of point $(x_1, y_1, z_1)$ satisfy

$$\frac{x_1}{a} = \frac{y_1}{b} = \frac{z_1}{c} = q_1 \Rightarrow \begin{cases} x_1 = aq_1 \\ y_1 = bq_1 \\ z_1 = cq_1 \end{cases}, R_1R.n_B = 0 \Rightarrow (x-x_1)a + (y-y_1)b + (z-z_1)c = 0. \tag{39}$$

So we have $q_1 = \frac{ax+by+cz}{a^2+b^2+c^2}$. Substitute $q_1$ into (39), we have $(x_1, y_1, z_1)$. The equation of line $R_1R_0'R$ is:

$$\frac{x_0' - x_1}{x - x_1} = \frac{y_0' - y_s}{y - y_1} = \frac{z_0' - z_s}{z - z_1} = q_2 \Rightarrow \begin{cases} x_0' = (x-x_1)q_2 + x_1 \\ y_0' = (y-y_1)q_2 + y_1 \\ z_0' = (z-z_1)q_2 + z_1 \end{cases}. \tag{40}$$

In fact, $q_2 = \frac{r}{\sqrt{(x-x_1)^2+(y-y_1)^2+(z-z_1)^2}}$. Substitute $q_2$ into Eq. (40), we get the coordinate $(x_0, y_0, z_0)$. To find $R_s' = (x_s', y_s', z_s')$, note that $R_0' - R_s'$ is parallel to $B_0$ so:

$$\frac{x_s' - x_0'}{a} = \frac{y_s' - y_0'}{b} = \frac{z_s' - z_0'}{c} = q_3 \Rightarrow \begin{cases} x_s' = aq_3 + x_0' \\ y_s' = bq_3 + y_0' \\ z_s' = cq_3 + z_0' \end{cases}. \tag{41}$$

Recall $z_s' = z_s \Rightarrow q_3 = \frac{z_s - z_0'}{c}$. Substitute $q_3$ into Eq. (41), we get the coordinate $R_s' = (x_s', y_s', z_s')$. Now it is possible to calculate the angle $\psi$ between $R - R_s'$ and $R_0' - R_s'$.

$$cos\psi = \frac{(R - R_s').(R_0' - R_s')}{|R - R_s'||R_0' - R_s'|} = \frac{(x - x_s')(x_0' - x_s') + (y - y_s')(y_0' - y_s') + (z - z_s')(z_0' - z_s')}{\sqrt{(x-x_s')^2 + (y-y_s')^2 + (z-z_s')^2}\sqrt{(x_0'-x_s')^2 + (y_0'-y_s')^2 + (z_0'-z_s')^2}} \tag{42}$$

Fig. 14 shows the results for a sheet emission source. As previous the ray tracing fit to the observed spectrum is quite good. This issue indicates that the emission source for the auroral-hiss emission could be a sheet source. The good fit also indicates that the trajectory of the spacecraft is nearly perpendicular to the cylindrical current sheet.

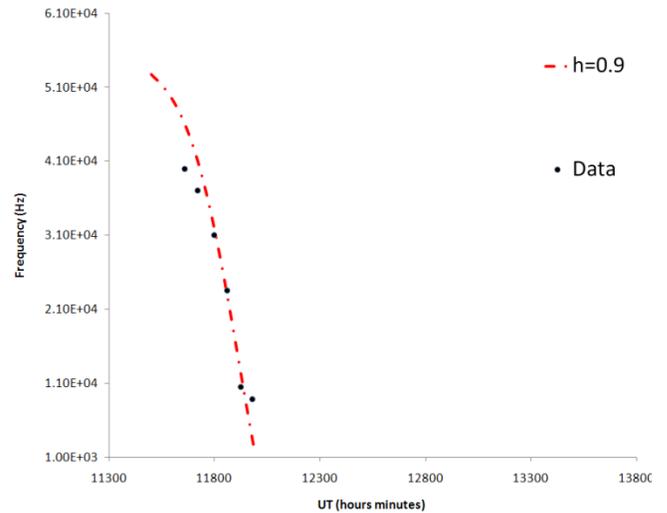

Figure 14. Cutoff Frequency vs. universal times for h=0.9R$_{IO}$ for sheet source model.

## III. Conclusion

A series of ray tracing computations have been performed by assuming a point and cylindrical sheet sources. It is found that the low-altitude boundary of the current source lies at near the equatorial plane of Io with coordinates $(-1.0434, 0.469, -0.16)$, which corresponds to a height of about 270 km from the surface of Io. From the electron density profile of the ionosphere of Io, it is possible to see that the current source well lies in the ionosphere of Io with a local electron density about $3 \times 10^4$ electrons per cubic centimeter.